\documentclass{aastex}
\usepackage{spr-astr-addons}

\RequirePackage{color}

\begin{document}
\title{Rayleigh-Taylor instability in an ionized medium}

\author{Mohsen Shadmehri\altaffilmark{1}}
\affil{School of Physics, Faculty of Science, Golestan University, Gorgan, Iran}
\and
\author{Asiyeh Yaghoobi\altaffilmark{2}} \and \author{Mahdi Khajavi\altaffilmark{2}}
\affil{School of Physics, Faculty of Science, Ferdowsi University, Mashhad, Iran}

\altaffiltext{1}{m.shadmehri@gu.ac.ir}

\begin{abstract}
We study  linear theory of the magnetized Rayleigh-Taylor instability  in a  system consisting of ions and neutrals. Both components are affected by
a uniform vertical gravitational field. We consider ions and neutrals as two separate fluid systems where they can exchange momentum through collisions.
However, ions have direct interaction with the magnetic field lines but neutrals are not affected by the field directly.  The equations of our two-fluid
model are linearized and by applying a set of proper boundary conditions, a general dispersion relation is derived for our two superposed fluids
separated by a horizontal boundary. We found two unstable modes for a range of the wavenumbers. It seems that one of the unstable modes corresponds to
the ions and the other one is for the neutrals. Both modes are reduced with increasing the collision rate of the particles and the ionization fraction. We show that if the two-fluid nature is considered, RT instability would not be suppressed and also show that the growth time of the perturbations increases. As an example, we apply our analysis to the Local clouds which seems to have arisen because of the RT instability. Assuming that the clouds are partially ionized, we find that the growth rate of these clouds increases  in comparison to a fully ionized case.
\end{abstract}



\section{introduction}
Rayleigh-Taylor (RT) instability occurs when a heavy fluid is supported by a lighter fluid in a gravitational field, or, equivalently,
when a heavy fluid is accelerated by a lighter fluid.  RT instability and the related processes have found applications in various astronomical
systems, such as the expansion of supernova remnants (e.g., Ribeyre et al. 2004) (where inertial acceleration plays the role of the gravitational
field), the interiors of red giants (e.g., Chairborne and Lagard 2010), the radio bubbles in galaxy clusters (Pizzolato and Soker 2006).
The evolution of the RT instability is influenced by many different factors. For example,  viscosity tends to reduce the growth rate and to stabilize
the system (e.g., Chandrasekhar 1961). Growth rate of the short-wavelength unstable perturbations decreases because of the  compressibility
(e.g., Shivamoggi 2008). A dynamically important radiation field affects RT instability as well (Jacquet and Krumholz 2011). However, the most important
effects in the astrophysical context are probably those due to the presence of the magnetic fields. One can decompose the magnetic field lines into a
component perpendicular to the interface and a component parallel to it. We will deal only with the effect of a tangential magnetic field.

Incompressible RT instability in a  plane parallel to a uniform tangential magnetic field  in both fluids
has been studied analytically by Chandrasekhar (1961). The linear stability theory shows that a tangential magnetic field slows down the growth rate
of the RT instability. The growth rate $\omega$ for the modes with wavenumber $k_x$ parallel to the magnetic field lines is given by
\begin{eqnarray} \omega^2=\frac{\rho_{02}-\rho_{01}}{\rho_{02}+\rho_{01}}gk_x-\frac{2B^2}{4\pi}\frac{k_x^2}{\rho_{02}+\rho_{01}}.
\end{eqnarray}
Here, we use Cartesian coordinates and denote the quantities of the plasma below the
discontinuity ($z\leq 0$) with a subscript 1 and those in above the discontinuity
($z \geq 0$) with a subscript 2. The magnetic field permeating the plasma is uniform and tangent to the discontinuity, so ${\bf B} = B {\bf e}_x$, while gravity is perpendicular to it, so ${\bf g} = -g{\bf e}_z$. If we set $B=0$,  the classical dispersion relation for RT instability is obtained, i.e.
\begin{eqnarray}
\omega^2=\frac{\rho_{02}-\rho_{01}}{\rho_{02}+\rho_{01}}g k_x.
\end{eqnarray}

There are a number of astrophysical systems in which the magnetic RT instability is expected to be important, among them are the  accretion onto the
magnetized compact objects (Wang and Nepveu 1983), buoyant bubbles generated by the radio jets in clusters of galaxies (Robinson et al. 2005), and the
thin shell of ejecta swept up by a pulsar wind (Bucciantini et al. 2004). But one should note magnetic RT growth rate (i.e., Equation (1)) is based on
the ideal MHD approximation, in which the multifluid nature is neglected for simplicity. However,  a  partially ionized plasma represents a state which
often exists. Thus, we are interested to  know how the growth rate of RT instability is modified in a partially ionized medium, in particular when the
coupling  between the ionized  and the neutral components is not complete. There are a few studies related to this issue. For example,
Chhajlani (1998) studied magnetic RT instability considering the surface tension and Finite Larmor Radius correction (FLR) in the absence of
gravity and the pressure gradient for the neutral particles. It was found that an increase in the collision frequency causes a decrease in the growth rate of the
system. Just recently, Diaz, Soler \& Ballester (2012) (hereafter DSB) studied RT instability in a partially ionized compressible (and also incomessible) plasma. Their purpose was to study the stability thresholds and the linear growth rate of the RT unstable modes in a two-fluid plasma consisting of ions and neutrals. They also studied the effects of compressibility and the collision between the particles. They calculated the growth rate as a function of the wavelength of the perturbations for  different values of the gravity and concluded that collisions are not able to fully suppress the instability. But in the highly collisional  regime the growth rate is significantly lowered by an order of magnitude in comparison to the classical result, specially for low values of gravity. Also, the linear growth rate is significantly lowered by compressibility and ion-neutral collisions compared to the incompressible collisionless case. Then, as an astrophysical implication, DSB applied their results to the solar prominences.

In this article, although we follow a similar problem to DSB , not only our presentation of the results are different from that study, but also we apply the results to another astronomical object which is subject to RT instability. More specifically, we study RT instability in a two-fluid magnetized medium consisting of the ions and the neutrals and obtain the growth rate of the unstable modes for different wavenumbers (not different gravity like DSB) and compare them with the fully ionized or neutral cases.
We also determine the most unstable mode and its relation to the wavenumber. Then, possible effects of ionization and the collisions on the growth rates are analyzed. More important, we apply our results to an astronomical object different from DSB system, i.e. the interaction zone between Loop1 and Local Bubble, which seems to be subject to RT instability and the maximum growth rate and the corresponding wavenumber are obtained.   Breitschwerdt, Freyberg and Egger (2000) showed the local clouds surrounding the solar system have been formed as a result of the growing magnetic RT instability  in the interaction zone between the Loop1 and Local Bubble. We now know that these clouds are partially ionized (Slavin  2008, welsh 2009). Thus, we can apply our results to the interaction region between the Loop I and the Local Bubble.  In the second section, we will present the
basic equations and the assumptions of the model. In the third section, we will analyze the growth rate of the unstable perturbations and finally in forth section we perform an application to interaction region between the Loop I and the Local Bubble.

\section{General Formulation}
We follow the analysis of Chandrasekhar (1961), but including neutrals and ions which are coupled via collisions.
Our basic equations are similar to the other related two-fluid studies (e.g., Shadmehri and Downes 2007),
but here there is a gravitational acceleration in the equations of motion for both the ions and the neutrals.  In
order to proceed analytically, it is assumed that the system is incompressible and the non-ideal dissipative process related to the evolution
of the magnetic field are neglected. So,  the convective term in the induction equation dominates the resistive one.
We suppose the neutral and ion components have not velocity in the unperturbed state. The magnetic field is assumed to be parallel to the
interface, i.e.  ${\bf B}=B_x{\bf e}_x $. Finally, all the unperturbed physics quantities are assumed to be spatially uniform in each medium.
The basic equations are
\begin{eqnarray}\label{eq:cont1}
\nabla\cdot{\bf u}_{n,i}=0,
\end{eqnarray}
\begin{eqnarray}\label{eq:cont2}
\rho_n\frac{D{\bf u}_n}{Dt}=-{\bf \nabla}p_n-\gamma_{i,n}\rho_n\rho_i({\bf u}_n-{\bf u}_i )-\rho_n g{\bf e}_z,
\end{eqnarray}
\begin{displaymath}\label{eq:cont3}
 \rho_i\frac{D{\bf u}_i}{Dt}=-{\bf \nabla}p_i-\gamma_{i,n}\rho_n\rho_i({\bf u}_i -{\bf u}_n )+
\end{displaymath}
\begin{eqnarray}\label{eq:cont4}
\qquad \qquad \qquad\qquad+\frac{1}{4\pi}(\nabla\times{\bf B})\times{\bf B}-\rho_i g{\bf e}_z,
\end{eqnarray}
\begin{equation}\label{eq:cont5}
 \frac{\partial {\bf B}}{\partial t} = \nabla \times ({\bf u}_i \times {\bf B}).
\end{equation}
\begin{equation}
\nabla.{\bf B} = 0,
\end{equation}
where $\gamma_{n,i}$ is the collision rate coefficients per unit mass so that $\nu=\gamma_{n,i}\rho_i$ is the neutral-ion collision frequency.
The collision frequency determines the  coupling between each component and the magnetic field. Here, ${\bf g}=-g {\bf e}_{z}$ is a uniform vertical gravitational acceleration. Now, we perturb the physical variables as $\chi(z,x,t)=\chi'(z)\exp(\omega t+ik_xx)$. Thus, linearized equations for the neutrals become
\begin{eqnarray}
 ik_x u'_n+\frac{\partial w'_n}{\partial z}=0,
\end{eqnarray}
\begin{eqnarray}
 \rho_n\omega u'_n=-i k_x p'_n-\gamma \rho_i\rho_n(u'_n-u'_i),
\end{eqnarray}
\begin{eqnarray}
 \rho_n\omega w'_n=-\frac{\partial p'_n}{\partial z}-\gamma \rho_i\rho_n(w'_n-w'_i),
\end{eqnarray}
where $u'_n$ and $w'_n$ are the $x$ and the $z$ components of the
perturbed velocity of the neutrals, respectively. Also, the linearized equations for the ions are
\begin{eqnarray}
  ik_x u'_i+\frac{\partial w'_i}{\partial z}=0,
\end{eqnarray}
\begin{eqnarray}
\rho_i\omega u'_i=-i k_x p'_i-\gamma \rho_i\rho_n(u'_i-u'_n),
\end{eqnarray}
\begin{eqnarray}
 \rho_i\omega w'_i=-\frac{\partial p'_i}{\partial z}-\gamma \rho_i\rho_n(w'_i-w'_n)+\nonumber
\end{eqnarray}
\begin{eqnarray}
 \qquad\qquad\qquad\qquad\qquad +\frac{B^2}{4\pi \omega}(\frac{\partial^2 w'_i}{\partial^2 z}-k_x^2w'_i),
\end{eqnarray}
where $u'_i$ and $w'_i$  are the $x$ and the $z$ components of the perturbed velocity
of the ions, respectively. After some mathematical manipulations, we can reduce the above differential equations to a set of two
differential equations for $w'_n$ and $w'_i$, i.e.
\begin{eqnarray}\label{8}
(\rho_i\omega+\frac{B^2k_x^2}{4\pi\omega})Dw'_i=\gamma\rho_i\rho_n D(w'_i-w'_n),
\end{eqnarray}
\begin{eqnarray}\label{9}
\rho_n\omega Dw'_n=\gamma\rho_i\rho_n D(w'_n-w'_i),
\end{eqnarray}
where $D=d^2/d^2z-k_x^2$. Up to this point, our basic linearized equations are similar to Shadmehri and Downes (2007) who studied two-fluid
Kelvin-Helmholtz  instability. However, the boundary conditions for RT instability are different from Kelvin-Helmholtz instability.
Having solutions of the above equations and by imposing  appropriate boundary conditions, we can obtain a dispersion relation for RT instability.

Behavior of the flow at the upper and the lower layers is determined by the general solutions of the linear differential equations
(\ref{8}) and (\ref{9}). One can easily show that the general solution of the equations is a  linear superposition  of two independent solutions
$\exp (+k_x z)$ and $\exp (-k_x z)$.  Now, we must apply the following proper boundary conditions to obtain a physical solution: (1) The perturbations tends to zero as $z$ goes to the infinity; (2) The  $z$-component  of the velocity is continuous at the interface; (3) The total pressure is also continuous at the interface. Thus, the general solutions become
\begin{eqnarray}\label{10}
w'_i=\left\{
\begin{array}{rl}
Ae^{+k_xz} &\qquad  {{\rm for}} \hspace{0.4cm} z < 0\\
A'e^{-k_xz} &\qquad  {{\rm for}}  \hspace{0.4cm}  z > 0,
\end{array} \right.
\end{eqnarray}
\begin{eqnarray}\label{11}
w'_n=\left\{
\begin{array}{rl}
Ce^{+k_xz} &\qquad  {{\rm for}} \hspace{0.4cm}  z < 0\\
C'e^{-k_xz} &\qquad  {{\rm for}} \hspace{0.4cm}  z > 0,
\end{array} \right.
\end{eqnarray}
where $C$, $C'$, $A$ and $A'$ are constants to be determined from the above boundary conditions. Continuity of the vertical displacement at $z=0$ (the second boundary condition) gives
the following relations
\begin{equation}
C=C'\qquad,\qquad A=A'.
\end{equation}
Also, based on the continuity of the ions and neutrals pressures at the interface $z=0$ (the third boundary condition), we have
\begin{eqnarray}
 (p'_i+p'_{i m}-\rho_i g\zeta_i)\big\vert_{0^+}=(p'_i+p'_{i m}-\rho_i g\zeta_i)\big\vert_{0^-},
\end{eqnarray}
\begin{eqnarray}
 (p'_n-\rho_n g\zeta_n)\big\vert_{0^+}=(p'_n-\rho_n g\zeta_n)\big\vert_{0^-},
\end{eqnarray}
where $p'_{im}$ is the perturbed magnetic pressure. Having  solutions (\ref{10}) and (\ref{11}), we can simply obtain  perturbed pressures and
substitute  them into the above equations. Therefore, we obtain
\begin{eqnarray}\label{eq:13}
&& \omega A(\rho_{i2}+\rho_{i1})+\gamma(A-C)(\rho_{i2}\rho_{n2}+\rho_{i1}\rho_{n1})+\nonumber\\
&& \qquad\qquad+\frac{2AB^2k_x^2}{4\pi\omega}-\frac{gA}{\omega}(\rho_{i2}-\rho_{i1})k_x=0,
\end{eqnarray}
\begin{eqnarray}\label{eq:14}
&& \omega C(\rho_{n2}+\rho_{n1})+\gamma (C-A)(\rho_{i2}\rho_{n2}+\rho_{i1}\rho_{n1})\nonumber\\
&& \qquad\qquad\qquad\qquad-\frac{gC}{\omega}(\rho_{n2}-\rho_{n1})k_x=0,
\end{eqnarray}
and after lengthy (but straightforward) mathematical manipulations, we then obtain
\begin{eqnarray*}
(\alpha_i+1)(\alpha_n+1)x^4+(\alpha_n\alpha_i+1)(\alpha_i+1+m \alpha_n+m)\lambda x^3
\end{eqnarray*}
\begin{eqnarray*}
 -\big[2y(\alpha_i\alpha_n-1)-2y^2(\alpha_n+1)\big]x^2+\big[2y^2\lambda(\alpha_i\alpha_n+1)-
\end{eqnarray*}
\begin{eqnarray*}
 -y(\alpha_i\alpha_n+1)\lambda\big(m\alpha_n-m+\alpha_i-1\big)\big]x -2y^3(\alpha_n-1)+\\
\end{eqnarray*}
\begin{equation}\label{eq:main}
\qquad\qquad\qquad\qquad+(\alpha_n-1)(\alpha_i-1)y^2=0,
\end{equation}
where
\begin{eqnarray*}
&& \alpha_i=\frac{\rho_{i2}}{\rho_{i1}},\alpha_n=\frac{\rho_{n2}}{\rho_{n1}}, m=\frac{\rho_{n1}}{\rho_{i1}},x=\frac{\omega}{g}v_A , y=\frac{k_x}{g}v_A^2
\end{eqnarray*}

\begin{equation}
\qquad\qquad\qquad \lambda=\frac{\nu}{g}v_A\qquad and\qquad (\nu=\gamma \rho_{i1}),
\end{equation}
 where $v_A$ is Alfven velocity for $z\leq0$. Dispersion relation  (\ref{eq:main}) is the main equation of our stability analysis. Obviously, if we neglect  collision between ions and neutrals (i.e., $\lambda=0$), equation (\ref{eq:main})
simply reduces to a dispersion relation for the ions which are tied to the magnetic field lines and another dispersion relation for the neutral component, i.e.
\begin{equation}
\Big(x^2(\alpha_i+1)+2y^2-(\alpha_i-1)y\Big)\Big(x^2(\alpha_n+1)-(\alpha_n-1)y\Big)=0.
\end{equation}
If we set the first parenthesis equal to zero, magnetic criterion for RT instability is obtained. Also, the second parenthesis gives the classical condition
of non-magnetic RT instability.

\section{Analysis}
Although dispersion relation for RT instability within one fluid approximation gives analytical solutions, it is very unlikely to obtain roots of equation
(\ref{eq:main}) analytically. So, we follow the problem numerically by assuming some numerical values for the input parameters.
 We assume $\alpha_{i}=\alpha_n=6$ and the dispersion relation (\ref{eq:main}) is numerically solved for different values
 for the parameters $\lambda$, $m$ and $y$. Obviously,  we would have an unstable mode if $x$ has a positive real part.

 Figure 1 shows non-dimensional growth rate $x=\frac{\omega}{g}v_A$ of the unstable modes versus the non-dimensional wavenumber $y=\frac{k_x}{g}v_A^2$.
 Parameter $m$ denotes the ratio  of densities of the neutrals and ions in layer 1. In Figure 1, we assume $m=1$ and each curve is labeled by the corresponding
 non-dimensional collision rate  $\lambda$. Figure 2 shows  growth
rate of the perturbations versus the wavenumber but for
$m = 100$. In this case, the ions are stable and the
neutrals determine  growth rate of the perturbations.
Also, Figure 3 and 4 show the growth rate of the neutrals
and the ions for $m = 0.01$, respectively. Again, we can see the
stabilizing role of the collision between the ions and
the neutrals.
 Note that when there is no collision between the ions and neutrals (i.e., $\lambda =0$), each component of our two-fluid system
 behaves independently. But when collision between ions and neutrals is considered, we found two unstable modes up to a certain wavenumber that are related on the
neutral and ion fluids separately, with the neutral ones having a tendency with much larger growth rate.
 Curves
 with the same color are corresponding to the same value of $\lambda$.
Since  magnetic field has a stabilizing role, we can see that unstable mode corresponding to the ions  has
a smaller growth rate in comparison to the neutrals unstable mode because of the coupling to the magnetic field lines. DSB obtained plots of dispersion relation of the unstable modes versus the
non-dimensional gravity but we plot unstable modes versus wavenumber. Therefor, we can conclude :\\
(1) The neutral mode is unstable for all wavenumbers but for ions we can find a critical wavenumber  for which the instability becomes ineffective. This result is valid irrespective of the ionization fraction and the collision rate.\\
(2) Growth rate of the unstable perturbations for the ions tends to become zero as the collision rate increases and thereby,  behavior of the system is determined by the growth rate
of the neutrals. Moreover, in this case, the profile of the growth rate for the neutrals is similar to the ions without collision. \\
(3)  However, as the collision frequency increases, not only growth rate of ions reduces but the unstable
perturbations for the neutrals are significantly reduced in particular at short wavelengths.\\
In the next section we apply our results to an astronomical object.


\section{Astrophysical Implication: interaction zone between Loop1 and Local Bubble }
 Our solar system is embedded in an ionized cloud named  Local Cloud. In vicinity
of Local Clouds there are also other cloudlets of comparable size. Winds and supernovae events that are associated
with clusters of massive early-type stars have a
profound effect on the surrounding interstellar medium
(ISM), including the creation of large cavities. These
cavities, which are often referred to as "interstellar
bubbles", are typically $\sim100 pc$ in diameter and
have low neutral gas densities of $n(H) \sim 0.01 cm^{-3}$ (Weaver et al. 1977). The local Clouds are inside a local X-ray emitting cavity which called the local bubble. Breitschwerdt et. al (2000) presented observational evidences based ROSAT PSPC data that manifest  existence of an interaction shell
between our local interstellar bubble and the adjacent Loop I
superbubble. They showed that due to the
overpressure in Loop I, a Rayleigh-Taylor instability
would operate, even in the presence of tangential magnetic field. Their calculations showed that the most unstable mode
has a growth time about $5\times 10^5$ years (depending
on magnetic field strength) which was in agreement with the interaction
time between the two bubbles. Moreover, the wavelength of the fastest growing mode was about 2.2 pc which was comparable to or less
than the thickness of the interaction zone. Also, Breitschwerdt et al. (2006) have performed 3D high resolution hydrodynamic simulations of the Local Bubble (LB) and the neighboring Loop I (L1) and reproduced the observed sizes of the Local
and Loop I superbubbles, the generation of blobs like the Local Cloud as a consequence of a dynamical instability (Breitschwerdt et al. 2006). Nevertheless, reports suggest the Local Clouds are partially ionized (e.g., Slavin 2008, Welsh et al. 2009). Since the clouds are formed due to RT instability (Breitschwerdt et al. 2000 and 2006), we conclude the interaction zone between Loop I and Local Bubble is partially ionized and because Loop I is hot, the interaction shell must be ionized. Now we are interested to know what effect does the ionization may have on the growth rate of system.  \\
We begin with Breitschwerdt and Slavin's assumptions for our system and obtain a dispersion equation from equations (\ref{eq:13})and (\ref{eq:14}) for a case where $\rho_1<<\rho_2 $ and $B_1<<B_2 $. Thus,
\begin{displaymath}
 \bigg(x^2+ mx\lambda+ y^2-y\bigg)\times\bigg(x^2 +\lambda x -y\bigg)
\end{displaymath}
\begin{equation}\label{appli}
-\lambda^2 mx^2=0,
\end{equation}
where
\begin{displaymath}
 m=\frac{\rho_{n2}}{\rho_{i2}},\qquad x=\frac{\omega}{g}v_{A2} , \qquad y=\frac{k_x}{g}v_{A2}^2,
\end{displaymath}
\begin{displaymath}
\qquad \lambda=\frac{\nu}{g}v_{A2}\qquad \nu=\gamma \rho_{i2},
\end{displaymath}
where $v_{A2}$ is Alfven velocity for $z\geq0$.
We have  the following input numbers (Breitschwerdt et al. 2000)
\begin{eqnarray}
&& \rho_2 =mn_{wall}=(2\times 10^{-24}gr)(10\times 0.5cm^{-3})=10^{-23}gr/cm^{-3}.\\
&& B_2=5\times 10^{-6}G.\\
&& g=\frac{8.7\times 10^{-12}gr/cms^2}{7.5 \times 10^{19}cm^{-2}\times 2\times 10^{-24}gr}=5.8\times 10^{-8}cm/s^2.
\end{eqnarray}

 Now, we can estimate $m$ and $\lambda$. Slavin (2008) found the ionization fraction is around $0.2-0.3$ for Hydrogen and around $0.3-0.4$ for Helium. Also, Welsh et al. (2009) presented  an amount of 0.1 for ionized fraction. If we suppose the ionized fraction is 0.1, then value of $m$ becomes 9. We can write the collision frequency (Shadmehri et al. 2008),
\begin{eqnarray}\label{5}
\nu=\frac{\sigma}{m_i+m_n}\rho _{i2},
\end{eqnarray}
where $\sigma=1.9\times10^{-9}cm^3s^{-1}$. We neglected ionization of Helium. Thus,
\begin{eqnarray*}
&& \lambda=\nu\frac{v_A}{g}=\nu\frac{B}{g\sqrt{4\pi\rho_{i2}}}=2923.
\end{eqnarray*}
Having the above input numbers, we can solve equation  ($\ref{appli}$) to find  growth time of the
fastest growing mode.  Figure 5 shows the results of such a calculation for the specified parameters. Then, the the most unstable mode of system  is
  \begin{eqnarray}
&&  x_{max}=1.58133=\frac{\omega_{max} v_A}{g}=\frac{\omega_{max} B}{g\sqrt{4\pi\rho_{i2}}}.\nonumber\\
\nonumber\\
&& \Longrightarrow\omega_{max} =1.58133\frac{g\sqrt{4\pi\rho_{i2}}}{B}=2.0558\times10^{-13}s^{-1}.\nonumber\\
&& \tau_*=\frac{1}{\omega_{max}}=1.54\times 10^{5}yrs.
\end{eqnarray}
We found shorter growth time in comparison to the classical magnetic RT instability (around $5\times10^5$yrs). The wavenumber of the fastest growing mode is
\begin{eqnarray}
&&  y_*=5=\frac{ k_*B^2}{g4\pi\rho _{i2}}.\nonumber\\
\nonumber\\
&& \Longrightarrow k_*=14.5696\times 10^{-19}cm^{-1}.\nonumber\\
&& d_*\sim\frac{1}{k_*}=6.86\times 10^{17}cm=0.22pc.
\end{eqnarray}
Thus, the size of structures formed by RT instability reduces in comparison to the classical magnetic RT instability that is $2.2pc$. When the nondimensional collision frequency lambda is large, the ionization fraction has also a vital role. Ionized particles are coupled to the magnetic field lines, but their coupling to the neutral particles is determined via collisions (i.e., $\lambda$). Now, we can consider two identical systems with the same input parameters except their ionization fractions. If both systems have a large collision rate, the system with a larger ionization fraction is more affected by the magnetic field lines in comparison to the system with a smaller ionization fraction. In other words, although the ionized and the neutral particles are coupled to the same level, but when the ionization fraction is larger the system is more under influence of the magnetic field lines. Figure 5 clearly shows this effect. Each curve is labeled by its ionization fraction. Here, the non-dimensional collision frequency is $\lambda = 2923$. We can see that as the ionization fraction increases, the growth rate of the unstable mode decreases simply because more particles are affected by the magnetic field lines.  It is because of the two-fluid nature of the system. The effect is more significant when the collision frequency decreases and the coupling between the ionized and the neutral particles is not complete. Thus, it seems that one-fluid approach is not adequate even when the collision frequency is large but the ionization fraction is not large enough. However, it is difficult to determine a critical value for lambda so that beyond which the system tends to MHD case. Because such a transition depends on the ionization fraction among the other input parameters. Moreover, when two-fluid approach is adopted the maximum growth rate is modified. But compressibility does not lead to such an effect. In fact, compressibility becomes less effective when the density contrast of the layers increases. So, the compressibility correction in a two-fluid system subject to RT instability depends on the density contrast of the layers.

\begin{figure}[h]
 \includegraphics[width=7cm,height=6cm]{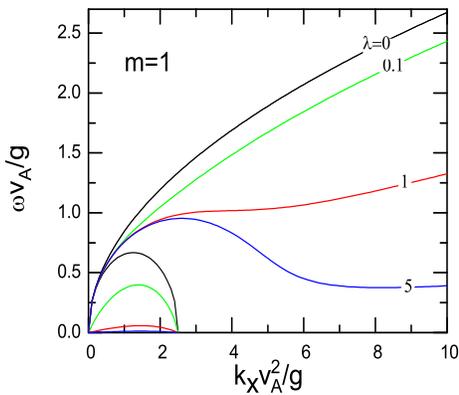}
 \caption{Growth rate of the RT instability versus the wavenumber of the perturbations  for $m=1$ and various non-dimensional collision rate
  $\lambda$. Curves with the same color are corresponding to the same value of $\lambda$.}
 \end{figure}
 \begin{figure}[ht]
\includegraphics[width=7cm,height=6cm]{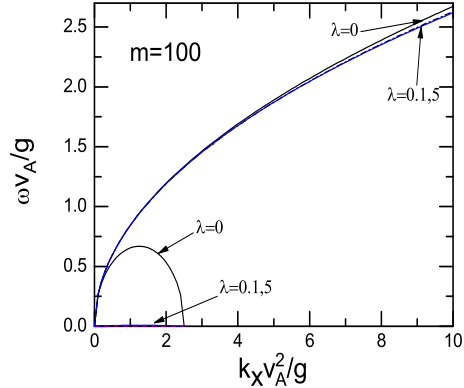}
 \caption{The same as Figure 1, but for $m=100$.}
 \end{figure}
  \begin{figure}[h]
 \includegraphics[width=7cm,height=6cm]{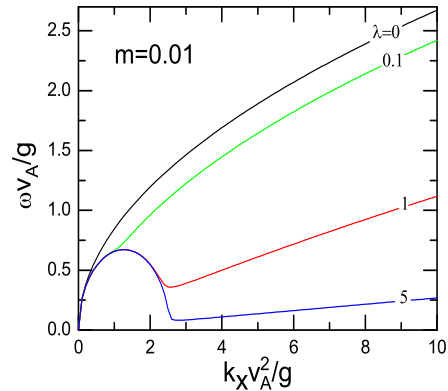}
 \caption{Growth rate of the RT instability for the neutrals versus wavenumber of the perturbations for $m=0.01$.}
 \end{figure}
  \begin{figure}[h]
 \includegraphics[width=7cm,height=6cm]{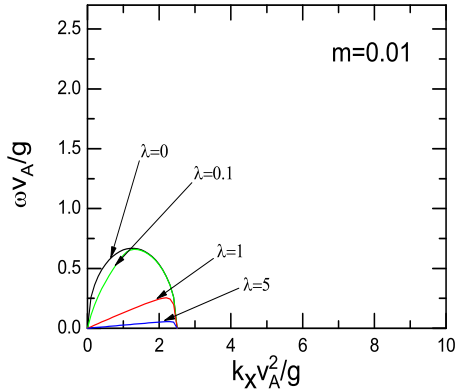}
  \caption{The same as Figure 3, but it shows growth rate for the ions.}
 \end{figure}
 \begin{figure}[hb]
 \includegraphics[width=7cm,height=6cm]{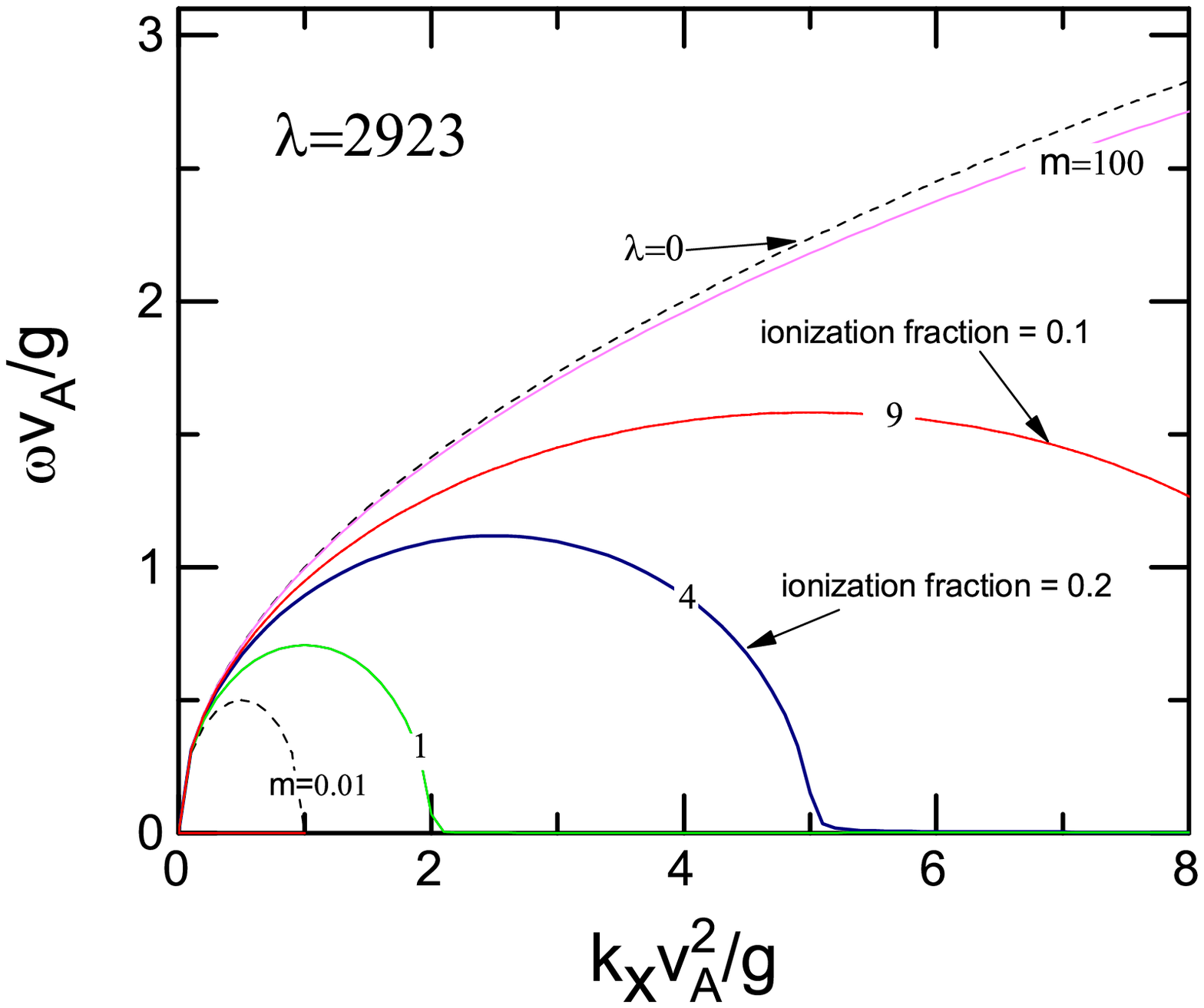}
 \caption{Growth rate of two-fluid RT instability in the interaction zone
 between Loop I and Local bubble, as a function of the
 wavenumber of the perturbation corresponding to  $\lambda=2923$ and different ionization fractions.}
 \end{figure}


\section{Conclusions}
In this study, we  investigated the magnetic RT instability in a two-fluid medium consisting of the neutrals and the ions. A general dispersion relation is obtained. By analyzing the unstable roots of the dispersion equation, two unstable modes are found  that are related to those of the neutral and ion fluids separately, with the neutral ones having a much higher growing rate. For each ionization fraction and collision rate, the stability of the system only depends on the behavior of the neutrals. For lower values of collision rate the curves are similar to the collisionless case. We found that the growth rate of the unstable perturbations decreases when the collision rate increases. Also, the instability of the system strongly depends on the ionization fraction. When the ionization fraction  increases, for a given collision rate, the growth rate of the perturbations decreases. Finally, we apply our results for interaction zone between the Loop I and the Local Bubble that is caused form local clouds. We obtained a shorter value for the growth time and size of the clouds. Although classical magnetic RT instability has been applied to this system for explaining some of the observed structures, our analysis shows that RT instability may
operates less effectively if the two-fluid nature of the system is considered. But we note that our results are valid within linear regime and non-linear numerical simulations are needed to confirm the linear results.

We note that magnetic effects only suppress the linear growth rate for perturbations aligned with the magnetic
field. Those perpendicular to it are unaffected. Indeed, as shown by Stone \& Gardiner (2007), in 3D the net effect of magnetic fields is actually to enhance the non-linear growth rate by suppressing secondary KH instabilities. Thus in the real world, it seems like magentic RT instability never occur. The hydrodynamic modes perpendicular to the field always end up taking over.

We also think that the Hall effect is an interesting problem, but it is beyond the scope of the present study. Our analysis is restricted to a two-fluid case, i.e. a system consisting of ion and neutral particles. We could also start from the one fluid MHD equations, but considering modified induction equation with resistivity, Hall and ambipolar terms. In that framework we could study possible effects of non-ideal terms (including Hall term). But we think it deserves a separate analysis independent of the present study.

{\bf Acknowledgment }

We are grateful to referees for their valuable  comments and suggestions which improved the paper.


\vspace{0.75cm}

{\bf REFERENCES}

\noindent Bucciantini N., Amato E., Bandiera R., Blondin J. M., Del Zanna L., 2004, A\& A, 423, 253

\noindent Breitschwerdt D., Freyberg M. J., Egger R., 2000, A\& A, 361, 303

\noindent Breitschwerdt D., de Avillez M., 2006, Astron. Astrophys.,452, L1

\noindent Chandrasekhar S., Hydrodynamic and hydromagnetic stability, International Series of Monographs on Physics, Oxford: Clarendon, 1961

\noindent Charbonnel C., Lagarde  N., 2010, A\& A, 522, A10

\noindent Diaz A. J., Soler  R., Ballester J. L., 2012, ApJ, 754, 41

\noindent Draine B. T., Roberge W. G., Dalgarno A., 1983, ApJ, 264, 485

\noindent Jacquet E., Krumholz M. R., 2011, ApJ, 703, 116

\noindent MacAlpine G. M., Satterfield T. J., 2008, ApJ, 136, 2152

\noindent Pizzolato F., Soker N., 2006, MNRAS, 371, 1835

\noindent Ribeyre X., Tikhonchuk  V. T., Bouquet  S., 2004, Physics of Fluids, 16, 4661

\noindent Robinson K., Dursi  L. J., Ricker P. M. and et al., 2004, ApJ, 601, 621

\noindent Shadmehri M., Downes T. P., 2007, Ap\& SS, 312, 79

\noindent Shadmehri M., Downes T. P., 2008, MNRAS, 387, 1318

\noindent Shadmehri M., Downes T. P., 2008, Asrton. soc., 387, 1318-1322.

\noindent Shivamoggi B. K., 2008, arXiv: 0805.0581

\noindent Slavin J. D., 2008, SSR, 143, 311-322.

\noindent Stone J. M., Gardiner T., 2007, ApJ, 671, 1726

\noindent Vaghela D. S., Chhajlani  R. K., 1988, Ap\& SS, 149, 301

\noindent Wang Y.-M., Nepveu M., 1983, A\& A, 118, 267

\noindent Weaver R., McCray R., Castor J., Shapiro P.,  Moore R., 1977, Astrophys. J., 218, 377

\noindent Welsh B.,Shelton R.,2009, Ap\& SS, 323, 1

\end{document}